\def\ts     {\thinspace}
\def\kms    {\ifmmode{{\rm \ts km\ts s}^{-1}}\else{\ts km\ts s$^{-1}$}\fi}
\def\msol   {\ifmmode{{\rm M}_{\odot} }\else{M$_{\odot}$}\fi}
\def\lsol   {\ifmmode{L_{\odot}}\else{$L_{\odot}$}\fi}
\def\lfir   {\ifmmode{L_{\rm FIR}}\else{$L_{\rm FIR}$}\fi}
\def\zsol   {\ifmmode{{\rm Z}_{\odot}}\else{Z$_{\odot}$}\fi}
\def\etal   {{\rm et\ts al.}}
\def\aco    {\ifmmode{{\rm CO}(J\!=\!1\! \to \!0)}\else{{\rm CO}($J$=1$\to$0)}\fi}
\def\taco    {\ifmmode{^{12}{\rm CO}(J\!=\!1\! \to \!0)}\else{$^{12}${\rm CO}($J$=1$\to$0)}\fi}
\def\bco    {\ifmmode{{\rm CO}(J\!=\!2\! \to \!1)}\else{{\rm CO}($J$=2$\to$1)}\fi}
\def\cco    {\ifmmode{{\rm CO}(J\!=\!3\! \to \!2)}\else{{\rm CO}($J$=3$\to$2)}\fi}
\def\dco    {\ifmmode{{\rm CO}(J\!=\!4\! \to \!3)}\else{{\rm CO}($J$=4$\to$3)}\fi}
\def\eco    {\ifmmode{{\rm CO}(J\!=\!5\! \to \!4)}\else{{\rm CO}($J$=5$\to$4)}\fi}
\def\fco    {\ifmmode{{\rm CO}(J\!=\!6\! \to \!5)}\else{{\rm CO}($J$=6$\to$5)}\fi}
\def\gco    {\ifmmode{{\rm CO}(J\!=\!7\! \to \!6)}\else{{\rm CO}($J$=7$\to$6)}\fi}
\def\hco    {\ifmmode{{\rm CO}(J\!=\!8\! \to \!7)}\else{{\rm CO}($J$=8$\to$7)}\fi}
\def\ico    {\ifmmode{{\rm CO}(J\!=\!9\! \to \!8)}\else{{\rm CO}($J$=9$\to$8)}\fi}
\def\jco    {\ifmmode{{\rm CO}(J\!=\!10\! \to \!9)}\else{{\rm CO}($J$=10$\to$9)}\fi}
\def\kco    {\ifmmode{{\rm CO}(J\!=\!11\! \to \!10)}\else{{\rm CO}($J$=11$\to$10)}\fi}
\def\pone  {\ifmmode{{\rm p-H}_2{\rm O}(1_{11}-0_{00})}\else{{p-H}$_2$O($1_{11}$-$0_{00}$)}\fi}
\def\ponep  {\ifmmode{{\rm o-H}_2{\rm O}^+(1_{11}-0_{00})}\else{{o-H}$_2$O$^+$($1_{11}$-$0_{00}$)}\fi}
\def\ptwo  {\ifmmode{{\rm p-H}_2{\rm O}(2_{02}-1_{11})}\else{{p-H}$_2$O($2_{02}$-$1_{11}$)}\fi}
\def\orone  {\ifmmode{{\rm ortho-H}_2{\rm O}(1_{10}-1_{01})}\else{{ortho-H}$_2$O($1_{10}$-$1_{01}$)}\fi}
\def\paone  {\ifmmode{{\rm para-H}_2{\rm O}(1_{11}-0_{00})}\else{{para-H}$_2$O($1_{11}$-$0_{00}$)}\fi}
\def\patwo  {\ifmmode{{\rm para-H}_2{\rm O}(2_{02}-1_{11})}\else{{para-H}$_2$O($2_{02}$-$1_{11}$)}\fi}
\def\paonep  {\ifmmode{{\rm ortho-H}_2{\rm O}^{+}(1_{11}-0_{00})}\else{{ortho-H}$_2$O$^{+}$($1_{11}$-$0_{00}$)}\fi}
\def\oone  {\ifmmode{{\rm o-H}_2{\rm O}(1_{10}-1_{01})}\else{{o-H}$_2$O($1_{10}$-$1_{01}$)}\fi}
\def\ci     {\ifmmode{{\rm C}{\rm \small I}}\else{C\ts {\scriptsize I}}\fi}
\def\hi     {\ifmmode{{\rm H}{\rm \small I}}\else{H\ts {\scriptsize I}}\fi}
\def\hh     {\ifmmode{{\rm H}_2}\else{H$_2$}\fi}
\def\cone {\ifmmode{{\rm C}{\rm \small I}(^3\!P_1\!\to^3\!P_0)}
     \else{C\ts {\scriptsize I}{\small$(^3\!P_1\!\to^3\!\!\!P_0)$}}\fi}
\def\ctwo {\ifmmode{{\rm C}{\rm \small I}(^3\!P_2\!\to^3\!P_1)}
     \else{C\ts {\scriptsize I}{\small$(^3\!P_2\!\to^3\!\!\!P_1)$}}\fi}
\def\cij {\ifmmode{{\rm C}{\rm \small I}\,(^3P_i\to^3P_j)}\else{C\ts {\scriptsize I}\,{\small$(^3P_i\to^3P_j)$}}\fi}
\def\cii    {\ifmmode{{\rm C}{\rm \small II}}\else{C\ts {\scriptsize II}}\fi}
\def\tex {\ifmmode{{T}_{\rm ex}}\else{$T_{\rm ex}$}\fi}
\def\tmb {\ifmmode{{T}_{\rm mb}}\else{$T_{\rm mb}$}\fi}
\def\tkin {\ifmmode{{T}_{\rm kin}}\else{$T_{\rm kin}$}\fi}
\def\microns {\ifmmode{\mu{\rm m}}\else{$\mu$m}\fi}
\def\nhh   {\ifmmode{n({\rm H}_2)}\else{$n$(H$_2$)}\fi}
\def\gradv {\ifmmode{{\rm dv/dr}}\else{dv/dr}\fi}
\documentclass[traditabstract]{aa} 
\usepackage{graphicx}
\usepackage{txfonts}
\begin{document}
   \title{HIFI spectroscopy of low-level water transitions in M82}
 \author{A. Wei\ss \inst{1}
        \and M.A. Requena-Torres \inst{1}
          \and R. G\"usten \inst{1}
          \and S. Garc\'ia-Burillo \inst{2}
	  \and A.I. Harris \inst{3}
          \and F.P. Israel \inst{4}
	  \and T. Klein \inst{1}
	  \and C. Kramer \inst{5}
	  \and S. Lord \inst{6}
          \and J. Martin-Pintado \inst{7}
          \and M. R\"ollig \inst{8}
	  \and J. Stutzki \inst{8}
          \and R. Szczerba \inst{9}
          \and P.P. van der Werf \inst{4}
          \and S. Philipp-May \inst{1}
          \and H. Yorke \inst{10}
	  \and M. Akyilmaz \inst{8}
          \and C. Gal \inst{8}
          \and R. Higgins \inst{13}
          \and A. Marston \inst{12}
          \and J. Roberts \inst{7}
          \and F. Schl\"oder \inst{8}
          \and M. Schultz \inst{8}
          \and D. Teyssier \inst{12}
          \and N. Whyborn \inst{11}
          \and H.J. Wunsch \inst{1}
          }

   \institute{Max-Planck-Institut f\"ur Radioastronomie, Auf dem H\"ugel 69, 53121 Bonn, Germany
         \and Observatorio Astronomico Nacional (OAN) - Observatorio de  Madrid, Alfonso XII 3, 28014 Madrid, Spain    
         \and Department of Astronomy, University of Maryland, College Park, MD 20742, USA
         \and Leiden Observatory, Leiden University, P.O. Box 9513, 2300 RA Leiden, The Netherlands 
         \and IRAM, Avenida Divina Pastora 7, 18012 Granada, Spain
         \and NASA Herschel Science Center, Caltech, Pasadena, CA, USA
         \and Centro de Astrobiologia (INTA-CSIC), Ctra de Torrejon a Ajalvir, km 4, 28850 Torrejon de Ardoz, Madrid, Spain 
         \and KOSMA, I. Physikalisches Institut der Universit\"at zu K\"oln, Z\"ulpicher Strasse 77, 50937 K\"oln, Germany 
	 \and N. Copernicus Astronomical Center, Rabia\'nska 8, 87-100 Toru\'n, Poland 
         \and Jet Propulsion Laboratory, 4800 Oak Grove Drive, Pasadena, California 91109, USA
         \and Atacama Large Millimeter/Submillimeter Array, Joint ALMA Office, Santiago, Chile
         \and European Space Astronomy Centre, ESA, P.O. Box 78, E-28691 Villanueva de la Ca\~nada, Madrid, Spain  
         \and Experimental Physics Dept., National University of Ireland Maynooth, Co. Kildare, Ireland 
          }

   \date{}

\abstract{We present observations of the rotational ortho-water ground transition, the two lowest para-water transitions, 
and the ground transition of ionised ortho-water in the archetypal starburst galaxy M82, performed 
with the HIFI instrument on the Herschel Space Observatory. These
observations are the first detections of 
the \paone\ (1113\,GHz) and \paonep\ (1115\,GHz) lines in an extragalactic source. 
All three water lines show different spectral line profiles, underlining the need for high 
spectral resolution in interpreting line formation processes. 
Using the line shape of the \paone\ and \paonep\ absorption profile in conjunction with high 
spatial resolution CO observations, we show that the (ionised) water absorption arises from a 
$\sim2000\,{\rm pc}^2$ region within the HIFI beam located about $\sim$50\,pc east of the dynamical 
centre of the galaxy. This region does not coincide with any of the known 
line emission peaks that have been identified in other molecular tracers, 
with the exception of HCO. Our data suggest that water and ionised water within this region have 
high (up to 75\%) area-covering factors of the underlying continuum. This indicates that water is 
not associated with small, dense cores within the ISM of M82 but arises from a more widespread diffuse gas component.}


   \maketitle
%

\section{Introduction}
High-resolution spectroscopy of far-infrared and sub-millimetre water lines is an important 
tool for studying the physical and chemical properties of the interstellar
medium (ISM). Absorption by terrestrial atmospheric water vapour has ground-based studies of water 
in extragalactic systems limited to radio maser transitions 
(such as the famous 22\,GHz water line) or to a few systems with significant redshift 
(e.g. Combes \& Wiklind \cite{combes97}, Cernicharo \etal\ \cite{pepe06}, Menten \etal\ \cite{menten08}). 
Earlier satellite missions, such as ODIN and SWAS,
did not have enough collecting area to detect the relatively faint ground transitions
of water in external galaxies. ISO and, more recently, {\em Spitzer} have provided the first systematic 
studies of water in the far infrared (IR) regime (e.g. Fischer \etal\ \cite{fischer99}, 
Gonzalez-Alfonso \cite{alfonso04}). These missions, however, did not cover the frequencies 
of the water ground transitions and other low-level water lines. 

Only with the launch of the Herschel\footnote{Herschel is an ESA space observatory with science instruments provided by 
European-led Principal Investigator consortia and with participation from NASA.} 
satellite, with its large
collecting area, have  these transitions 
become accessible in the nearby universe (e.g. van der Werf \etal\ \cite{paul10}). 
As part of the HEXGAL guaranteed time key program (PI\, G\"usten), we are surveying the low-level water 
lines in different nuclear environments, performing velocity-resolved spectroscopy with the 
Heterodyne Instrument for the Far Infrared (HIFI, de Graauw \etal\ \cite{deGraauw}). 
In this letter we report on our first observations towards the central region of the 
archetypal starburst galaxy M82. We adopt a distance of 3.9\,Mpc (Sakai \& Madore \cite{sakai99}).
  

\section{Observations and data reduction}
Using the HIFI instrument onboard Herschel, we have observed the ground transitions of
ortho and para water, \oone\ and \pone, as well as the \ptwo\ line,
towards the centre of M82 (RA=$09^h55^m52^s.22$ Dec=$69^\circ40'46''.9$ J2000).
Observations were carried out in fast-chopping dual-beamswitch mode using 
a wobbler throw of $3'$ for all observations. The wobbler frequency 
was 0.8, 1.4, and 2.0 Hz for the 557\,GHz, 988\,GHz, and 1113\,GHz
observations, respectively. Nodding was performed every $\sim40$ seconds. 
Calibration was achieved through hot/cold absorber measurements every 20 minutes.
The data were recorded using the wide-band acousto-optical
spectrometer, consisting of four units with a bandwidth of 1\,GHz each, covering 
the 4\,GHz IF for each polarization with spectral resolution of 1\,MHz.

\begin{figure*} \centering
\includegraphics[width=17.0cm]{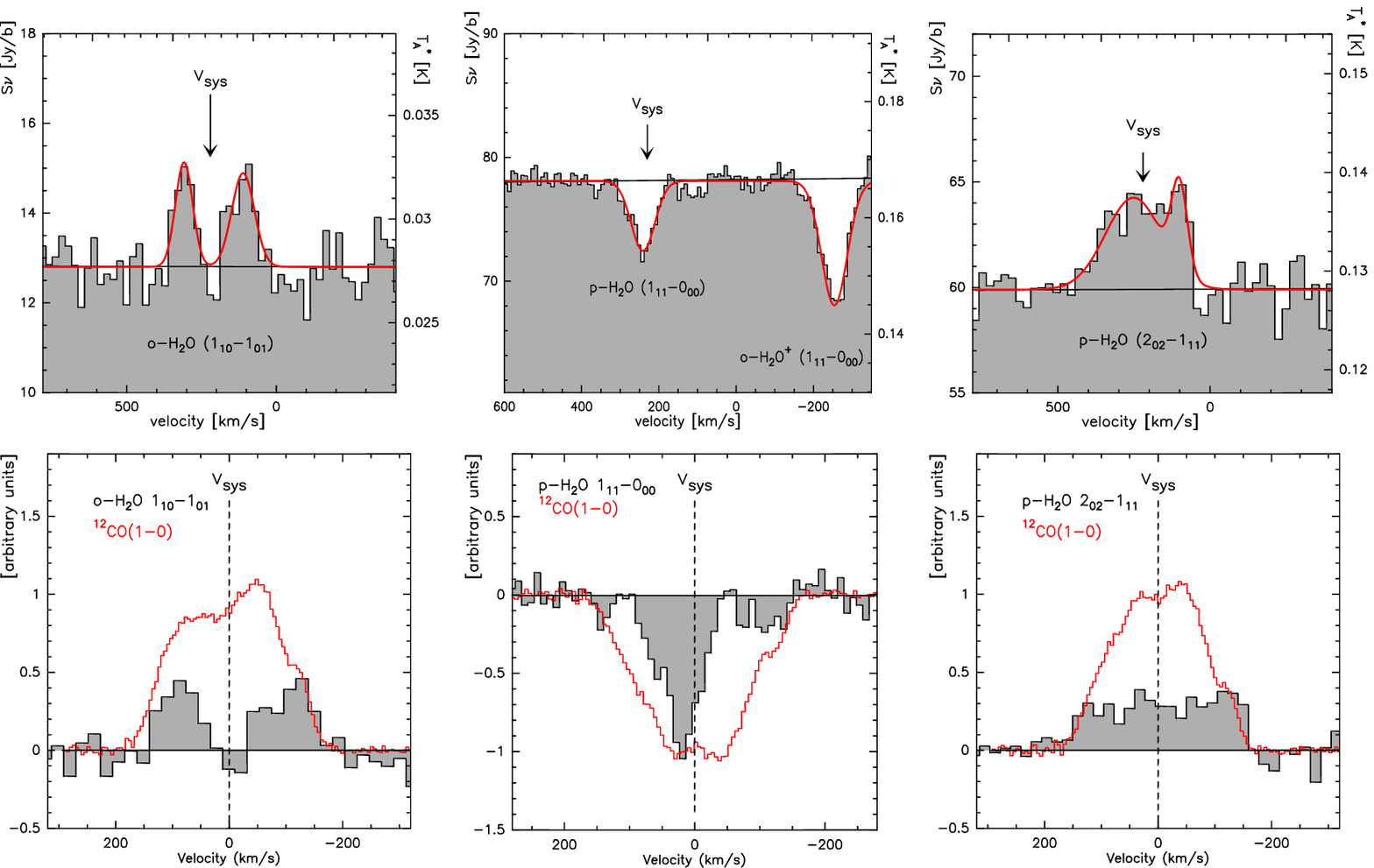}
\caption{{\it Top:} Spectra of the \oone, \pone\ and \ptwo\ lines 
 towards the centre of M82, with Gaussian-fit profiles superposed. The line parameters are 
given in Table \ref{linepara}. {\it Bottom:} Water-line profiles superposed on the \aco\ line
profiles within beams synthesized with the same spatial resolution as the water data. The velocity
scale is relative to the systemic velocity of M82 ($v_{\rm LSR}=225$\,\kms). The intensities of 
the CO line profiles have been normalized to unity, and the water line profiles scaled to 
provide close matches to the CO profiles in the line wings. The CO profile in the 
middle panel has been inverted for better comparison to the water absorption profile.}  
\label{lines} 
\end{figure*}

Data were reduced using the HIPE\footnote{Herschel Interactive Processing Environment} 
and CLASS\footnote{http://www.iram.fr/IRAMFR/GILDAS}  
software packages. Spectra were calibrated using HIPE and then exported
to CLASS format with the shortest possible pre-integration (typically $\sim40$ seconds). 
For each scan we combined the four sub-bands in each polarization to
create a 4\,GHz spectrum. From this spectrum we computed the underlying continuum using the line-free
channels; we then subtracted first-order baselines from individual sub-bands. The baseline-subtracted 
sub-bands were again combined and the continuum level added. This results in a
noise-weighted 4\,GHz spectrum for each scan. These spectra were inspected 
for remaining baseline instabilities, and scans with distorted baselines were omitted. 
For each line we inspected the co-added result in both polarizations (H and V) separately.
The continuum level was found to agree better than 10\% for each frequency. The 
integrated line intensities agree within 25\% without significant differences of the line
profiles (although the latter comparison is limited by the signal-to-noise ratio, in particular
for the 557\,GHz observations). In the following we therefore use the noise-weighted average of 
both polarizations, which yields effective on-source integration times for the three lines are 750, 2250, and 1300 seconds
for the \oone, \pone, and \ptwo\ lines, respectively.


Since HIFI's calibration is still preliminary, we used the theoretical 
predictions based on the Herschel's expected surface accuracy and the 
geometrical aperture size to convert the antenna temperatures to flux 
density (Kramer \cite{kramer06}). This yields 462 Jy\,K$^{-1}$, 467 Jy\,K$^{-1}$, and 
470 Jy\,K$^{-1}$ at  557\,GHz, 988\,GHz, and 1113\,GHz, respectively. The final
spectra are shown at a velocity resolution of 20, 10, and 25 \kms\ in Fig.\,\ref{lines} (top).

\section{Results}

All three water lines have been detected with high significance, demonstrating
that faint (few mK) broad lines can be observed with HIFI.
The \oone\ line is detected in emission and shows a double-peaked line profile. 
Within noise uncertainty, no emission (or
absorption) is detected on the systemic velocity of M82 (v$_{\rm LSR}=225$\,\kms). 
Both components are well fit by Gaussian profiles, with line parameters given in 
Table \ref{linepara}. We detect a continuum level of 12.8 Jy\,beam$^{-1}$ at 557\,GHz.

Our 1113\,GHz spectrum shows two absorption features, one centered at the systemic
velocity of M82 corresponding to the \pone\ line, and a second 
stronger feature blue-shifted by $\sim1.8$\,GHz with respect to the systemic velocity.
We identify the blue-shifted absorption feature as \ponep\ 
($\nu_{\rm rest}=1115.186$\,GHz, M\"urtz \etal\ \cite{muertz98}). Both absorption line
profiles are identical within the uncertainties (see Fig. \ref{absprofiles}).
The absorption profile is approximated reasonably well by a single Gaussian 
(see Table \ref{linepara} for the line parameters). The continuum level detected
at 1113\,GHz is 78 Jy\,beam$^{-1}$.

The \ptwo\ line at 988\,GHz is detected in emission. Its line profile differs
from both the \oone\ emission profile and the \pone\ absorption profiles.
It shows emission between v$_{\rm LSR}=50-500$ \kms\ (similar to the velocity range
covered by CO at the same spatial resolution, see below) and is almost flat-topped
for velocities between 100 and 400\,\kms. Thus the spectrum does not indicate 
absorption (or absence of emission) at the systemic velocity. The
spectrum can be decomposed into two Gaussian profiles, with parameters given in 
 Table \ref{linepara}. The continuum flux detected at 988\, GHz is 61 Jy\,beam$^{-1}$.

\begin{table*}
\caption[]{Line parameters derived from a Gaussian fit.}
\label{linepara}
\centering
\begin{tabular}{l c c r c c c c c}
\hline\hline
\noalign{\smallskip}
Line & $\nu_{\rm rest}$ & beam size& T$^{*}_{\rm A}$ peak & S$_{\nu}$ peak & \multicolumn{1}{c}{I$_{\nu}$} & v$_{\rm LSR}$ & dv &  S$_{\nu}$ cont.\\
     &[GHz] & [$''$] &[mK] & [Jy beam$^{-1}$] & \multicolumn{1}{c}{[Jy beam$^{-1}$\,\kms]} & [\kms] & [\kms]&  [Jy beam$^{-1}$] \\
\noalign{\smallskip}
\hline
\noalign{\smallskip}
\oone & 556.936 & 41 & $4.5 \pm 1.1$ & $2.1\pm0.5$ & $200\pm33$ & $110\pm8$ & $90\pm15$&  12.8 (540\,$\mu$m) \\
      &   &  & $5.0 \pm 1.1$ & $2.3\pm0.5$ & $170\pm29$ & $308\pm6$ & $68\pm11$&   \\[0.5ex]
\ptwo & 987.927 & 23 & $8.9 \pm 2.0$ & $4.2\pm0.9$ & $273\pm90$ & $100\pm9$ & $61\pm20$&  60.0 (300\,$\mu$m) \\
      &   &          & $9.3 \pm 2.0$ & $4.4\pm0.9$ & $1005\pm130$ & $250\pm15$ & $220\pm27$&   \\[0.5ex]
\pone & 1113.343 & 20 & $-12.0 \pm 1.6$ & $-5.6\pm0.7$ & $-432\pm28$ & $240\pm2$ & $72\pm6$&  78.1 (270\,$\mu$m) \\[1ex]
\ponep & 1115.186 & 20 & $-21.4 \pm 1.6$ & $-10.1\pm0.7$ & $-836\pm29$ & $242\pm2$ & $77\pm5$&   \\
\noalign{\smallskip}
\hline

\end{tabular}
\end{table*}

\section{Discussion}

\subsection{The line profiles \label{sect-profile}}
\begin{figure} \centering
\includegraphics[width=7.0cm]{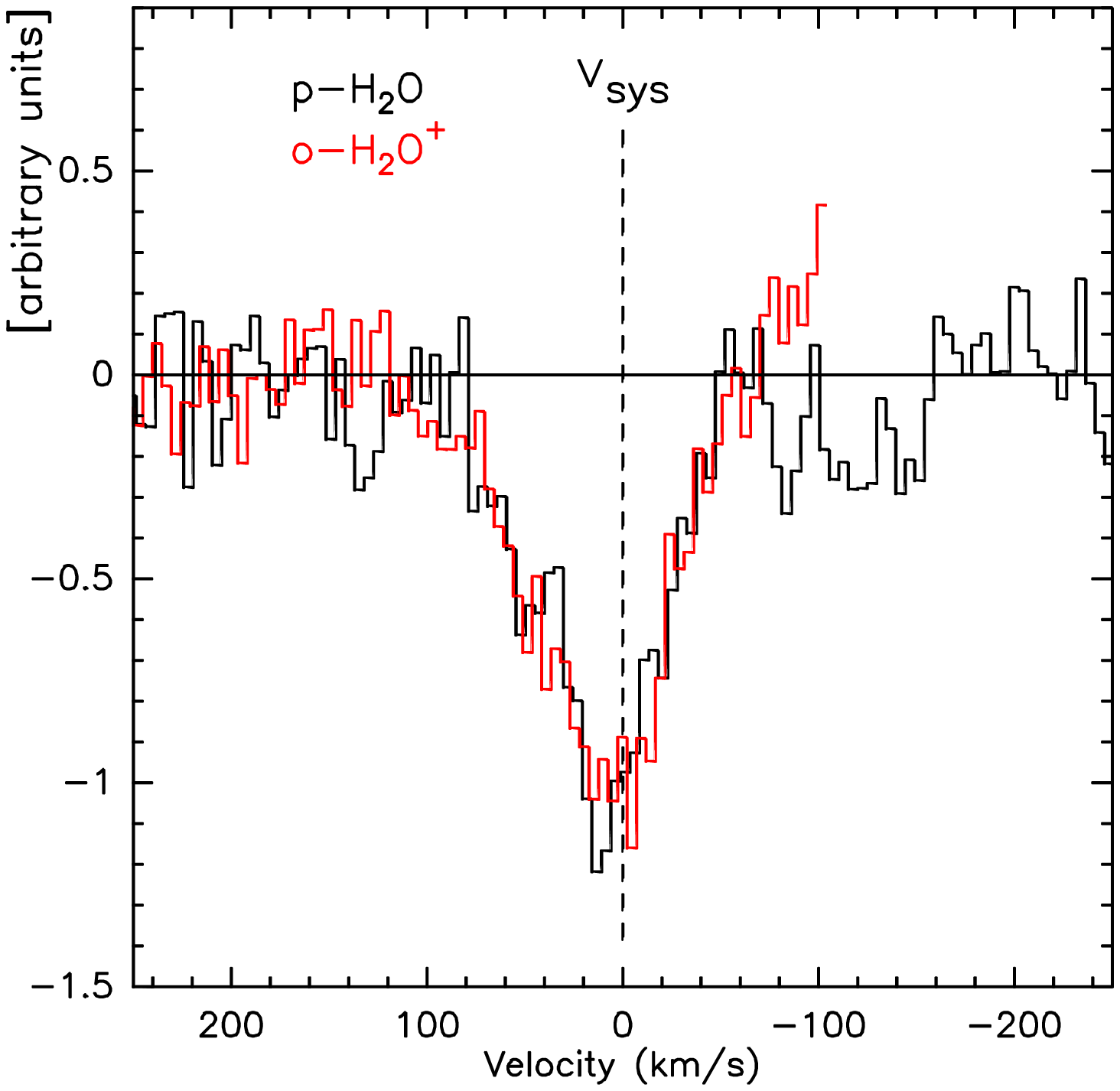}
\caption{Line profiles of the \pone\ (black) and the \ponep\ (red) lines. The velocity
scale is relative to the systemic velocity of M82. Both profiles have been normalized to
a maximum absorption of unity.}  
\label{absprofiles} 
\end{figure}

Given the large body of high spatial resolution observations of molecular gas
tracers published for M82, the line profile of the water lines can be
compared to other data to learn more about the location and extent of
the water emitting/absorbing regions in the disk. We here compare the
water line profiles to the high spatial resolution ($3.5''$) 
\taco\ data cube obtained by Walter \etal\ (\cite{walter02}); $^{12}$CO 
(CO thereafter) is the best-studied tracer of the molecular
gas in M82.  We first compare the CO spectra in beams synthesized to
the same spatial resolutions as the HIFI beams. From the comparison of
the \pone\ absorption spectrum to CO, it is apparent that the
absorption is not only detected in the pronounced absorption feature
close to the systemic velocity, but also at velocities in the wings of
the CO profile (Fig. \ref{lines} bottom).  It is therefore tempting to
speculate that the lack of absorption at certain velocities has a
geometrical origin, i.e., that gas at these velocities is located
behind the continuum.  This is also supported by the shape of the
\ptwo\ line emission profile, which shows that water is abundant in
the gas phase of M82 at all velocities where CO (i.e. molecular gas)
is present.  The very good correspondence of the \ponep\ absorption
profile near the systemic velocity suggests that the ionised water
traces the same gas as is detected in the water absorption. A closer
inspection of the \ponep\ profile shows, however, a lack of absorption in the
red wing of the line profile (see Fig.\ref{absprofiles}). The
blue wing is only partly covered by our spectrum and shows emission at
the very blue edge. This could come from calibration uncertainties at
the edge of the IF-band, but it is unclear whether this feature is an
artifact or is real. \\
Our finding that the \pone\ line is observed in absorption while the 
\ptwo\ line is detected in emission can be used to obtain an estimate
of the excitation temperatures of both lines. We used the dust model by Siebenmorgen \& 
Kr\"ugel (\cite{siebenmorgen07}) to estimate a background temperature of 18\,K and 20\,K at 988\, 
and 1113\,GHz, respectively. This implies T$_{\rm ex} < 20$\,K for \pone\ and T$_{\rm ex} > 18$\,K
for \ptwo\ (or T$_{\rm ex}\approx 19$\,K if both lines are close to LTE). Given the complexity of 
the water energy level diagram and the various level population channels (collisional or radiative), 
detailed models will be required for investigating the underlying excitation mechanisms.

\begin{figure} \centering
\includegraphics[width=8.7cm]{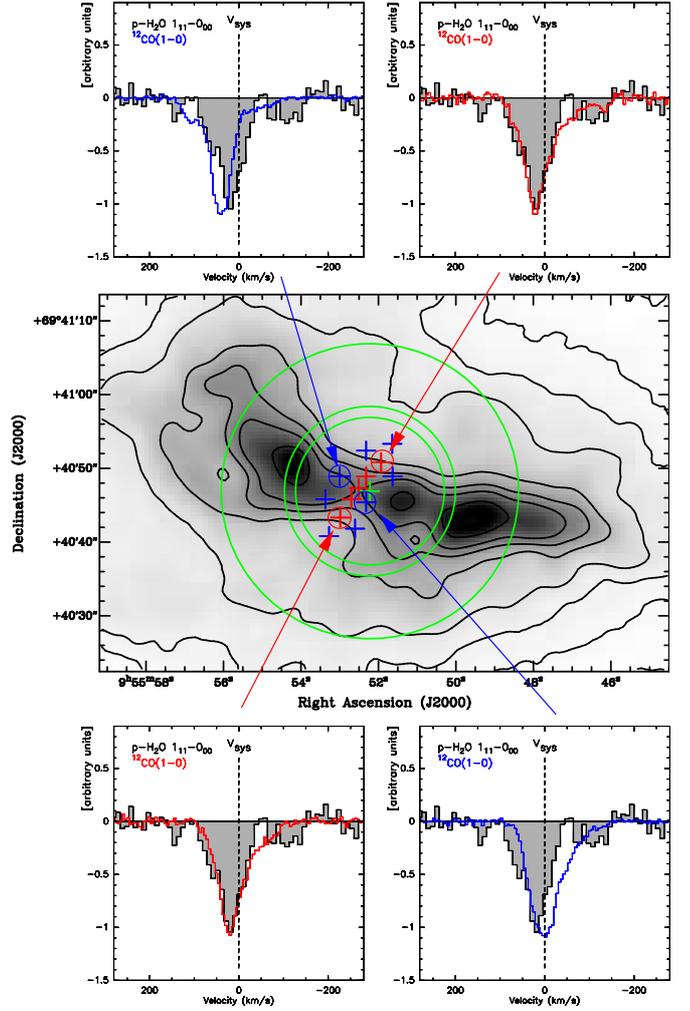}
\caption{HIFI beam sizes (FWHM) for our observing frequencies shown on the integrated \aco\ 
distribution (Walter \etal\ \cite{walter02}) in the central region of M82. The green cross
indicates our pointing centre which corresponds to the dynamical centre of M\,82. The red crosses 
indicate the locations where the CO profiles at $3.5''$ resolution match
the \pone\ absorption profiles, and the blue crosses indicate locations where the CO profile 
significantly differs. The spectra to the top and bottom show example CO spectra superposed 
on the water absorption profiles. Their positions have also been marked by a circle for 
better visualization.}
\label{co-profilematch} 
\end{figure}

Owing to the  much  larger  beam  size  of  the  \oone\ observations,   a
comparison  to  the  other  water lines  is   not  straightforward. In
comparison to CO,  however, it is  apparent  that the \oone\  emission
arises  exclusively from  velocities in  the   line wings  of  the  CO
spectrum.   These  velocities  mainly  correspond  to emission  in the
southwestern and  northeastern   molecular  lobes that are   located
within  the  $41''$  beam  (see  Fig.\ref{co-profilematch}  for the CO
distribution compared to the HIFI beams), and yet we  cannot rule out that
the lack of  \oone\ emission near  the  systemic velocities is due  to
self absorption  in the water line profile.  Observations of the water
lines within  the molecular lobes  will be required to  obtain similar
spatial   coverage of all  lines,  thus  allowing detailed analysis of
different  line profiles and modeling  of the water excitation.  These
observations have been approved and will be presented in a forthcoming
paper.

We retrieved sub-beam location information by comparing CO line
profiles at the highest spatial resolution ($3.5''$) to the \pone\ and
\ponep\ line profiles. The results are displayed in Fig.\,\ref{co-profilematch},
where we show selected CO spectra overlaid on the \pone\ absorption
profile. We find that the CO line profile is in very good agreement
with the water absorption profile in only a small region within the
HIFI beam. The region delineated in Fig.\,\ref{co-profilematch} 
corresponds to a small strip
orthogonal to the molecular disk of M82. The CO profiles east and west of
this region show significant shifts of their line centroids compared
to the peak of the water absorption due to the rotation of the
molecular disk. Interestingly, the CO line profiles that match the
shape of the main water absorption profile also show a blue wing. This
CO emission corresponds to molecular gas in the outflow of M82 (Walter
\etal\ \cite{walter02}). This shows that the water absorption in the
blue line wing arises not only from the northwestern molecular lobe
(which has similar velocities and is at least partly covered by the
HIFI beam), but also from gas in the outflow.

\subsection{The origin of the H$_2$O/H$_2$O$^+$ absorption}
We used the observed absorption depth to investigate whether the \pone\ and \ponep\ 
absorption arises from small cores or from a more widespread gas phase of the ISM. For this
we assume that both lines are optically thick. This implies that the absorption depth directly
measures the continuum covering factor, once the continuum distribution is known. We here use the
3\,mm continuum shown in Fig.\ref{compare_hco} as a proxy for the submm continuum distribution.
Although the continuum at this wavelength is not automatically a good indicator of the thermal
dust emission, because of possible free-free contamination, the spatial similarity between the
the 3\,mm and radio continua at high spatial resolution, as well as the similarity of 
the 3\,mm if smoothed to similar resolution to the published 350\,$\mu$m maps (Leeuw \& Robson \cite{leeuw09}), 
justifies this approach. The total continuum emission of M82 at 1113\,GHz was estimated using
the dust model by Siebenmorgen \& Kr\"ugel (\cite{siebenmorgen07}). Combining this with the 
observed 3\,mm continuum distribution, we estimate $S_{\rm 1113\,GHz} = 13.5$\,Jy within the region of 
the water absorption. This translates into a continuum covering factor of 40\% for \pone\ and 75\% for
\ponep. This approach yields predicted continuum fluxes within the HIFI beams that 
agree within 10\% or better with the observed continuum fluxes listed in Table\,\ref{linepara}
at all three frequencies. Since the water absorption region covers $\sim2000\,{\rm pc}^2$, these
numbers strongly suggest that the \pone\ and \ponep\ absorption arise from a widespread gas phase
of the ISM and not from dense cores (which are expected to cover a much smaller fraction of the 
continuum).. 
Considering the high critical densities for collisional excitation
of water (a few times $10^7$\,cm$^{-3}$, Faure \etal\ \cite{faure07}), this also implies that the water 
excitation cannot be driven by collisions, but is most likely dominated
by the IR field. This conclusion agrees with findings in other IR bright galaxies such
as Arp220 and Mrk231 (e.g. Gonzalez-Alfonso \etal\ \cite{alfonso04}, Gonzalez-Alfonso \etal\ \cite{alfonso10}).

We can also use the absorption depth to calculated the absorbing gas column densities assuming
that the absorption is not intrinsically saturated and using the continuum strength derived above (13.5 Jy). 
Applying the method by Menten \etal\ (\cite{menten08}) we derive $\int \tau_{app} dv$ of 38.6\,\kms\ and 
107.2\,\kms\ from the \pone\ and \ponep\ absorption profiles. With the low-excitation temperature approximation,
$N_l=\frac{8\pi\nu^3}{A_{ul}c^3}\frac{g_l}{g_u}\, \int \tau_{app} dv$, this yields ground-level column densities 
of $N(p$-${\rm H}_2{\rm O}\,000)=9.0\,\cdot 10^{13}\,{\rm cm}^{-2}$  and 
$N(o$-${\rm H}_2{\rm O}^+\,000)=2.2\,\cdot 10^{14}\,{\rm cm}^{-2}$. For the latter, the fine structure
splitting of $o$-${\rm H}_2{\rm O}^+$  (see e.g. Ossenkopf \etal\ \cite{volker10}) has been taken into account.
Assuming an ortho-to-para ratio of 3:1 this results in lower limits for the total column densities of 
$N(\rm H_2O)\ge3.6\,\cdot 10^{14}\,{\rm cm}^{-2}$ and $N(\rm H_2O^+)\ge2.9\,\cdot 10^{14}\,{\rm cm}^{-2}$. 
By converting the measured CO intensities in the absorption region to proton column densities 
using an $X_{\rm CO}$ conversion factor as given in Walter \etal\ (\cite{walter02}), we derive 
limits for the fractional abundances of [${\rm H_2O}$]$\ge4.0\cdot10^{-9}$ and [${\rm H_2O^+}$]$\ge3.3\cdot10^{-9}$.
These numbers are 2-3 magnitudes lower than the water abundances derived for Arp\,220 and Mrk\,231 
($\sim10^{-6}$, Gonzalez-Alfonso \etal\ \cite{alfonso04}, Gonzalez-Alfonso \etal\ \cite{alfonso10}), 
but approach values found in molecular cloud cores, the diffuse gas in the Milky Way, and in the $z=0.89$ absorption
system PKS\,1830-211 (e.g. Snell \etal\ \cite{snell00}, van der Tak \etal\ \cite{vdT10}, Menten \etal (\cite{menten08}).

\subsection{Comparison to other molecular gas tracers}

\begin{figure} \centering
\includegraphics[width=9.0cm]{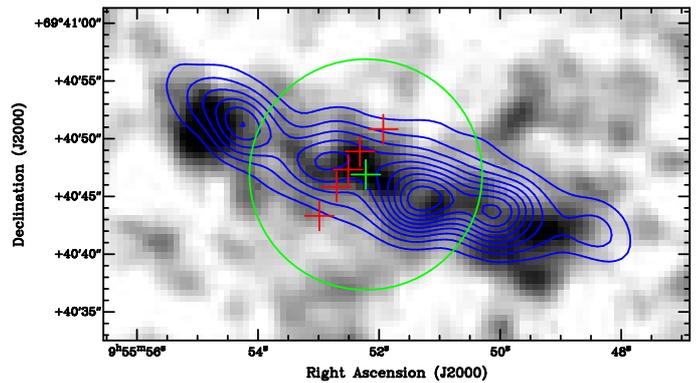}
\caption{3mm continuum emission (blue contours, Wei\ss\ \etal\ \cite{weiss01}) superposed on the 
integrated HCO distribution (grayscale, Garc\'ia-Burillo \etal\ \cite{garcia02}). The green circle 
indicates the FWHM of the HIFI beam at the \pone\ frequency, the red crosses indicate the region 
of the water absorption as derived based on the CO line profiles (see Fig.\ref{co-profilematch}).
}
\label{compare_hco} 
\end{figure}

As shown in Fig.\,\ref{co-profilematch} the water absorption 
region is not associated with a particular CO-bright region
in M82, but is located between the northeastern molecular lobe
and the central CO peak. This indicates that the water absorption
is not associated with the bulk of the molecular gas in M82 or that
the underlying continuum distribution favors an absorption in regions 
that are not bright in CO. The latter can be ruled out by comparing 
the spatial distribution of the water absorption to the radio (Wills \etal\ \cite{wills97}) 
and mm-continuum (Wei\ss\ \etal\ \cite{weiss01}) as a proxy for
the spatial distribution of the submm continuum (e.g. using the FIR-radio correlation).
Neither tracer provides evidence that the submm continuum is in particular pronounced
towards  the region of the water absorption (see Fig.\,\ref{compare_hco}), which suggests
that the small size of the water absorption is related to the ISM properties rather than
to the continuum distribution. 

We have further compared the spatial distribution of the dense gas
as traced by HCN (Brouillet \& Schilke \cite{brouillet93}) and H$^{13}$CO$^+$ 
(Garc\'ia-Burillo \etal\ \cite{garcia02}) to the water absorption region. These
tracers, however, closely follow the distribution of the CO emission, which rules
out any interpretation that the water absorption occurs predominately in high-density regions. 

A possible mechanism to efficiently release water from dust grains
into the gas phase are shocks (e.g. Cernicharo \etal\
\cite{pepe99}). A comparison to the SiO distribution in M82
(Garc\'ia-Burillo \etal\ \cite{garcia01}) shows that there is no
indication of strong shocks in this particular region. The SiO
observations, however, do not rule out the possibility that weaker
C-type shocks could release H$_2$O efficiently from the dust grains
into the gas phase without disrupting the grains themselves. In this
context it is interesting to note that the water absorption is located
near the expected location of the x2 orbits (and their intersection
with the x1 orbits) proposed in the context of gas motions in the
stellar bar potential of M\,82 (e.g. Greve \etal\ \cite{greve02}).
Shocks associated with the orbital intersection would provide a
plausible mechanism for releasing water into the gas phase, and the
geometry of the orbits would naturally explain why the absorption is
only detected to the east of the dynamical centre of M\,82 (see
Fig.\,2 in Greve \etal\ \cite{greve02}).

On the other hand molecular abundances in M82 are thought to be largely influenced
by chemistry in photon-dominated regions (PDRs, see e.g. Martin \etal\ \cite{martin06}).
UV photons photo-dissociate water and PDR models predict that significant water 
gas phase abundances should only be present in well-shielded, high column density regions 
(see e.g. Meijerink \& Spaans \cite{meijer05}). This interpretation is challenged by
our finding that the water absorption region in M82 is located in a relatively low column density
region within the disk of M82. Furthermore the water absorption coincides 
with a peak of the spatial distribution of HCO (Garc\'ia-Burillo \etal\ \cite{garcia02},
see Fig.\ref{compare_hco}).  HCO is a tracer of PDRs, and it is the only molecule so far imaged at 
high spatial resolution that shows enhanced emission in this region. The
similarity of the \pone\ and \ponep\ absorption profiles shows that ionizing photons are 
present in the absorbing medium. Therefore UV dissociation could suppress the water gas phase abundances, leading
to much lower water abundances in M82 compared to those derived in Arp\,220 and Mrk\,231. 
This is in line with studies of other molecules with low ionization potential, such as as NH$_3$ and HNCO, 
for which photo dissociation has been suggested as a main driver for the low abundances observed in M82 
(Wei\ss\ \etal\ \cite{weiss01b}, Martin \etal\ \cite{martin09}).

\begin{acknowledgements}
HIFI has been designed and built by a consortium of 
institutes and university departments from across Europe, Canada, and the 
United States under the leadership of SRON Netherlands Institute for Space 
Research, Groningen, The Netherlands, and with major contributions from 
Germany, France, and the US. Consortium members are: Canada: CSA, 
U.Waterloo; France: CESR, LAB, LERMA, IRAM; Germany: KOSMA, 
MPIfR, MPS; Ireland: NUI Maynooth; Italy: ASI, IFSI-INAF, Osservatorio 
Astrofisico di Arcetri - INAF; Netherlands: SRON, TUD; Poland: CAMK, 
CBK; Spain: Observatorio Astronmico Nacional (IGN), Centro de Astrobiologia 
(CSIC-INTA). Sweden: Chalmers University of Technology - MC2, RSS \& 
GARD; Onsala Space Observatory; Swedish National Space Board, Stockholm 
University - Stockholm Observatory; Switzerland: ETH Zurich, FHNW; USA: 
Caltech, JPL, NHSC.\\
AH, SL acknowledge support for this work by NASA through an award issued by JPL/Caltech. 
JMP and JR have been partially supported by MCINN grant ESP2007- 65812-CO2-01.
RSz acknowledges support from grant N 203 393334 from the Polish MNiSW. 
\end{acknowledgements}

\end{document}